\documentclass[a4paper]{spie}  

\usepackage{comment}
\usepackage{amsmath,amsfonts,amssymb}
\usepackage{graphicx}
\usepackage[table,xcdraw]{xcolor}
\usepackage[normalem]{ulem}
\useunder{\uline}{\ul}{}
\usepackage{amsfonts}
\usepackage{tabularx}
\usepackage{svg}
\usepackage{subcaption}
\usepackage{pdfpages}
\usepackage[colorlinks=true, allcolors=blue]{hyperref}

\title{TempoRL: laser pulse temporal shape optimization with Deep Reinforcement Learning}

\author[2]{F. Capuano}
\author[1]{D. Peceli}
\author[2]{G. Tiboni} 
\author[2]{R. Camoriano}
\author[1]{B. Rus}
\affil[1]{ELI Beamlines, Dolní Břežany, Czechia}
\affil[2]{VANDAL Laboratory, DAUIN, Politecnico di Torino, Turin, Italy}

\authorinfo{
Francesco Capuano (corr. author): \texttt{francesco.capuano@studenti.polito.it} \\
Gabriele Tiboni, Raffaello Camoriano: \texttt{\{gabriele.tiboni, raffaello.camoriano\}@polito.it} \\
Davorin Peceli, Bedrich Rus: \texttt{\{davorin.peceli, bedrich.rus\}@eli-beams.eu}
}
\begin{document} 
\maketitle

\begin{abstract}
High Power Laser's (HPL) optimal performance is essential for the success of a wide variety of experimental tasks related to light-matter interactions.
Traditionally, HPL parameters are optimized in an automated fashion relying on black-box numerical methods.
However, these can be demanding in terms of computational resources and usually disregard transient and complex dynamics.
Model-free Deep Reinforcement Learning (DRL) offers a promising alternative framework for optimizing HPL performance since it allows to tune the control parameters as a function of system states subject to nonlinear temporal dynamics without requiring an explicit dynamics model of those. Furthermore, DRL aims to find an optimal control policy rather than a static parameter configuration, particularly suitable for dynamic processes involving sequential decision making.
This is particularly relevant as laser systems are typically characterized by dynamic rather than static traits. Hence the need for a strategy to choose the control applied based on the current context instead of one single optimal control configuration.
This paper investigates the potential of DRL in improving the efficiency and safety of HPL control systems.
We apply this technique to optimize the temporal profile of laser pulses in the L1 pump laser hosted at the ELI Beamlines facility. 
We show how to adapt DRL to the setting of spectral phase control by solely tuning dispersion coefficients of the spectral phase and reaching pulses similar to transform limited with full-width at half-maximum (FWHM) of $\sim$1.6 ps. 
The code base of this work, alongside a live demo of the results obtained, is available at~\href{https://www.github.com/fracapuano/TempoRL}{github.com/fracapuano/TempoRL}.
\end{abstract}


\section{INTRODUCTION}\label{sec:intro}

Ultra-fast light-matter interactions such as laser-plasma physics and nonlinear optics require precise shaping and exact knowledge of the pulse temporal profile. 
Optimization of  laser pulse temporal shape is one of the most critical tasks necessary to establish control over these interactions. 
The highest intensities conveyed by laser pulses are usually achieved by compressing a pulse to its shortest possible duration, \textit{i.e.}, the transform-limited (TL) pulse shape, given the spectral intensity profile. 
However, some interactions may require unique pulse shapes different from the TL profile to achieve specific desired outcomes and the best efficiency while at the same time protecting the system from  potential damage. 
In this work, we explore a novel approach based on Deep Reinforcement Learning (DRL) to control ultra-fast laser pulse shape through an active feedback loop between the pulse shaper and the pulse duration measurement device.

The high-power laser system L1 Allegra has been developed at ELI Beamlines in the Czech Republic~\cite{Batysta:14}. 
It is designed to deliver $<20$ femtoseconds pulses with energy higher than 100 mJ at a repetition rate of 1 kHz. 
The L1 laser contains 7 optical parametric chirped pulse amplification (OPCPA) stages that are pumped by 5 diode-pumped lasers based on commercial Yb-doped thin-disk regenerative amplifiers (RA) DIRA 200-1 (by TRUMPF Scientific lasers). 
All pump lasers generate pulses at 1030 nm with a 1 kHz repetition rate. Before amplification, pump pulses are stretched to about 500 ps and then compressed again before frequency doubling to 515 nm in LBO crystals for second harmonic generation (SHG).
The SHG output efficiency of DIRA 200-1 can significantly improve by tuning the laser pulse temporal shape. Optimizing the temporal pulse shape to achieve the best SHG output is accomplished by manipulating the pulse spectral phase imposed by a CFBG fiber Stretcher (by Teraxion) and monitored using single-shot SH FROG 1030 (by Femtoeasy).

In the L1 system, pulse spectral phase control can be performed by separately tuning three dispersion coefficients on the stretcher ---namely $d_2$, $d_3$, and $d_4$--- which are linearly related to the second, third, and fourth term of the Taylor expansion of the spectral phase, $\varphi(\omega)$, around its central frequency $\omega_0$.
In other words, one could these dispersion coefficients to influence group delay dispersion ($\text{GDD} \sim  \varphi^{(2)}(\omega)$), third order dispersion ($\text{TOD} \sim \varphi^{(3)}(\omega)$) and fourth-order dispersion ($\text{FOD} \sim \varphi^{(4)}(\omega)$).
In the scope of this work, the control parameters for the L1 system are these three Taylor coefficients only (GDD, TOD, FOD), which we collectively refer to with $\psi$.

\subsection{Related works}
Several works have already investigated active feedback optimization of ultra-fast laser systems parameters or laser-matter interactions via black-box optimization methods~\cite{baumert1997femtosecond, yelin1997adaptive, arteaga2014supercontinuum, woodward2016towards, dann2019laser, loughran2023automated, shalloo2020automation}. 
The most common automated approaches are based on Bayesian Optimization (BO)~\cite{loughran2023automated, shalloo2020automation} and Genetic Algorithms (GA)~\cite{baumert1997femtosecond, arteaga2014supercontinuum, woodward2016towards}.
Recent studies by Loughran et al.~\cite{loughran2023automated} and Shalloo et al.~\cite{shalloo2020automation} highlight the limitations of performing a simple 1D grid search (often referred to with the term \textit{scan}) on the available values for each parameter, even when  scans are conducted in the vicinity of promising configurations. 
These studies suggest that such an approach can lead to sub-optimal solutions, as it fails to account for the joint effect that various parameters are likely to have on the overall system performance.

In particular, Loughran et al.~\cite{loughran2023automated} tackle proton energy maximization by controlling six laser parameters.
In parallel to initially performing grid scans in promising regions, BO is employed to optimize the same parameters simultaneously.
BO is shown to achieve better results than grid scans in a fraction of the time, likely due to BO's ability to consider the joint effect of the parameters.
At the same time, convergence speed is strongly related to the way in which BO optimizes laser performance. 
By probing points mainly in regions judged as promising in light of an underlying surrogate model of the objective function, BO can discard other parts of the parameter space that grid scans would explore exhaustively instead.
In their work, energy maximization is formulated as a 6D optimization problem related to tuning five Zernike mode coefficients and the tape surface position relative to the focal plane of the laser, employing online diagnostic tools to adjust the laser parameters using a closed feedback loop. 
Upon convergence, this method yields a feasible solution producing proton beams with maximum energy equivalent to manually-optimized pulses while using 60\% of the actual laser energy.

Loughran's BO-based approach to maximum proton energy mainly stems from the work of Shalloo et al.~\cite{shalloo2020automation}, which applies BO to automate tuning of a 100 MeV-scale accelerator, simultaneously varying up to six parameters, including the spectral and spatial phase of the laser and the plasma density and length. 
The authors observe that tuning the laser pulse shape with BO causes an 80\% increase in measured electron beam charge, despite the pulse length changing by just 1\%. 
Both works~\cite{loughran2023automated, shalloo2020automation} show how BO can be employed in a closed loop to automatically tune laser parameters within a feasible number of function evaluations, given a scalar objective function, such as peak intensity. 

In a previous paper~\cite{capuano2022laser}, we explored the application of BO and other numerical methods, such as GA and gradient-based optimization, to the problem of controlling the temporal pulse shape by only changing the spectral phase of laser pulses. 
The performance of our past approach was tested using a so-called “semi-physical” model of the L1 pump laser system, modeling the main stages of the actual pump chain. 
The laser pulse shape was optimized through a feedback loop between the stretcher and various measurement devices, which we used to estimate the similarity between the controlled and target temporal pulse shapes.
Note that in Ref.~\citenum{capuano2022laser}, we adopted the transform-limited as our target pulse shape. 

\subsection{Current limitations}
\begin{figure}[]
    \centering
    \begin{minipage}[]{0.45\textwidth}
        \centering
        \includegraphics[width=\textwidth]{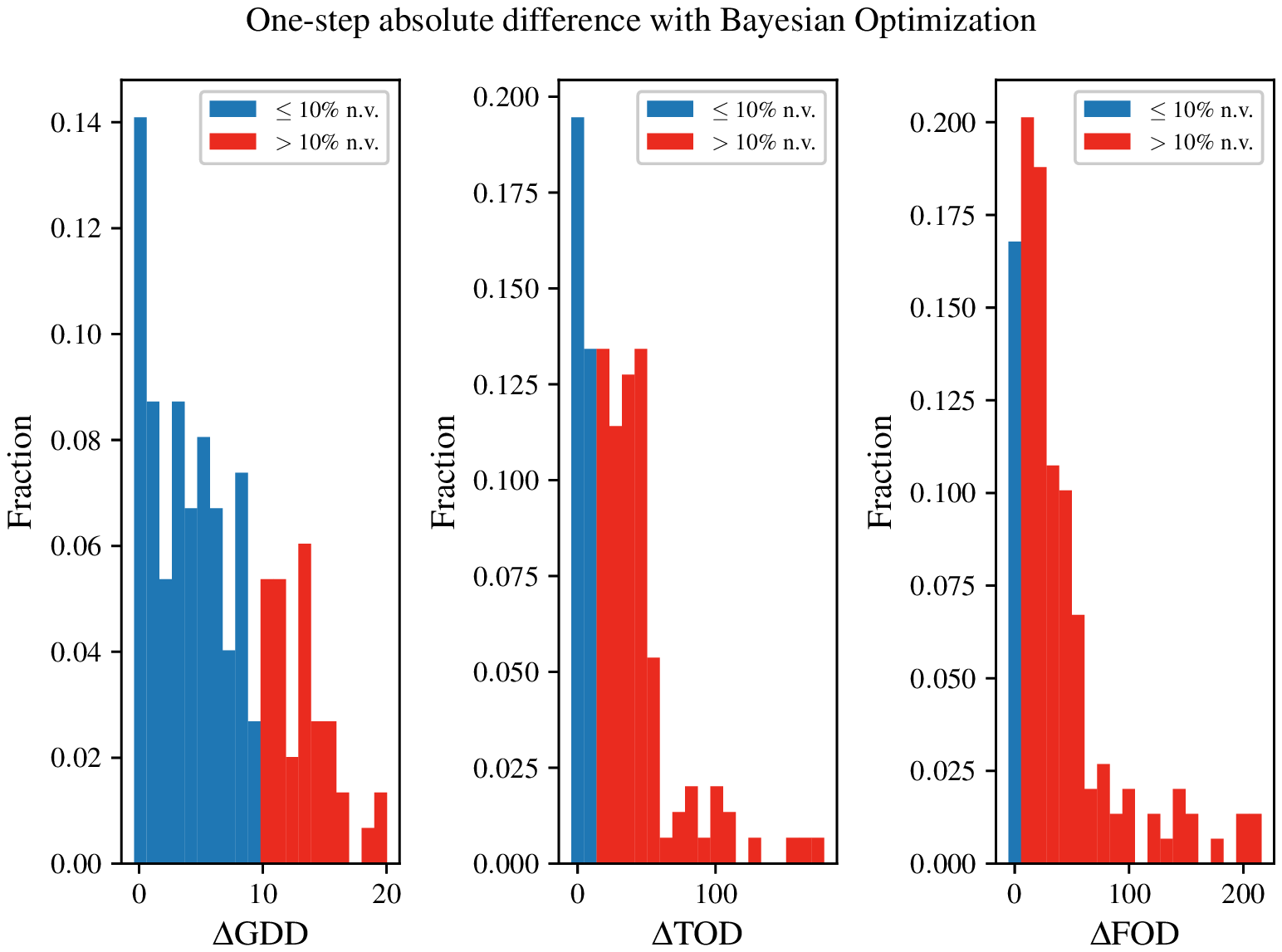}
        \caption{One-step ($k-1 \rightarrow k$) absolute percentage change in control parameters $\big(100 \cdot \frac{\vert \psi_i(k) - \psi_i(k-1) \vert }{\psi_i(k-1)}\big)$. Plots are specific to the actual control parameter. One-step percentage differences that are over 10\% are colored in red.}
        \label{fig:BayesControls}
    \end{minipage}
    \hfill
    \begin{minipage}[]{0.45\textwidth}
        \centering
        \includegraphics[width=\textwidth]{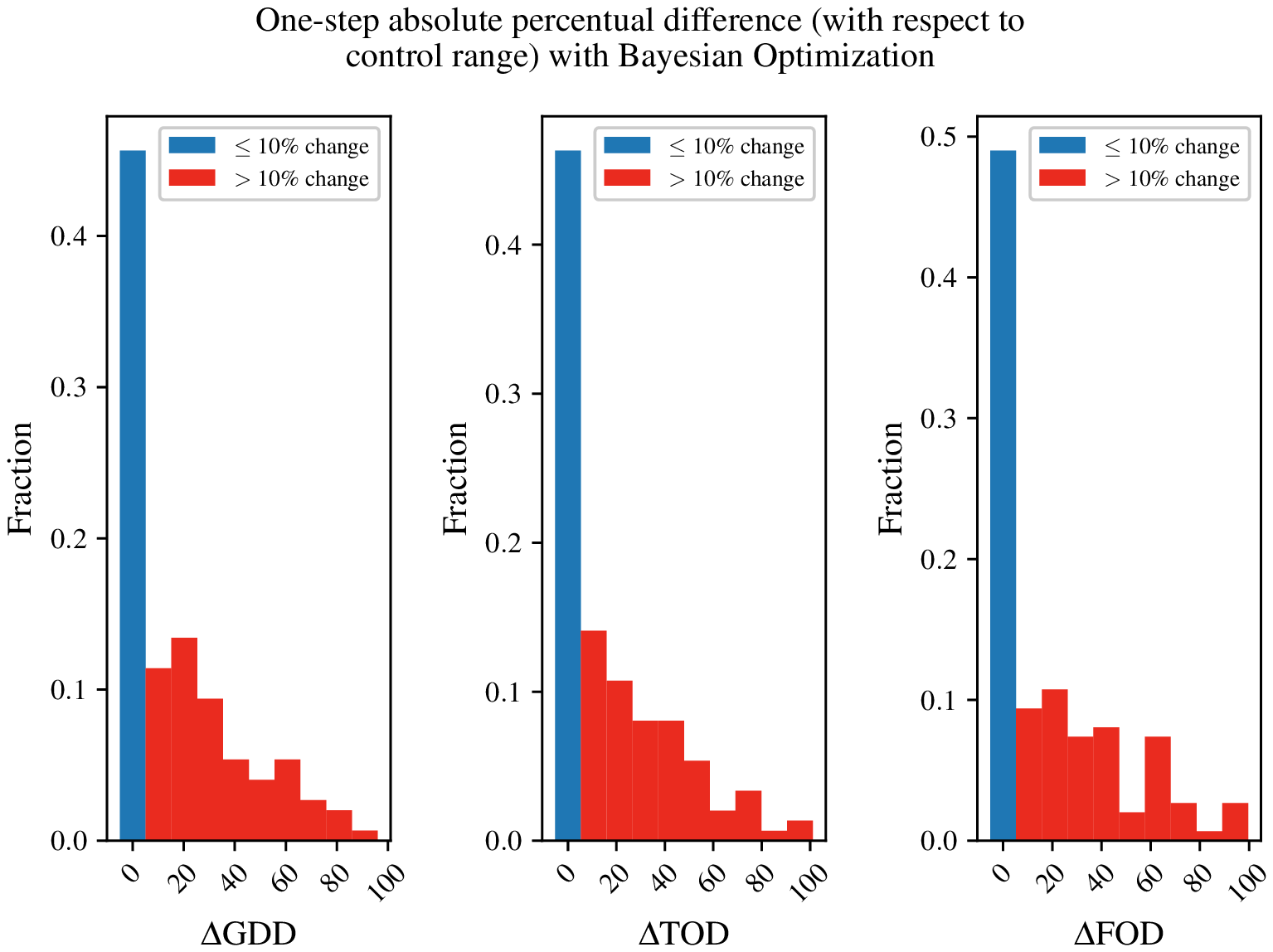}
        \caption{One-step absolute percentual change in control parameters with respect to physical control range $\big(100 \cdot \frac{\vert \psi_i(k) - \psi_i(k-1) \vert }{\psi^{\max}_i - \psi^{\min}_i}\big)$. Plots are specific to the actual control parameter. One-step percentual differences that are over 10\% are colored in red}
        \label{fig:BayesControls_ActionRange}
    \end{minipage}
\end{figure}

Since in the real world, each function evaluation is highly expensive---as it requires an actual laser burst for a given $\psi$---we are interested in favoring approaches achieving good results with few evaluations. 
Out of the three algorithms implemented in Ref. \citenum{capuano2022laser}, BO requires the least number of function evaluations.
Both GA  and gradient-based (ADAGRAD) optimization require a significantly larger number of experimental samples. 
As each individual's fitness must be computed for the GA to be able to favor the fittest ones, this algorithm typically requires significantly more function evaluations than the number of iterations the algorithm is allowed to run. 
Furthermore, ADAGRAD employs first-order differential information that, for black-box problems, can only be estimated using finite differences instead of actual derivatives, thus introducing non-negligible analytical errors and the cost associated with increasing the number of function evaluations to be performed at every timestep.
Nonetheless, the number of laser shots to be performed before the algorithm converges is not necessarily the only complication of using numerical methods to optimize laser pulse temporal shape. 
During the initial exploration phase (\textit{i.e., }as long as the surrogate model has not well approximated the underlying objective function), BO needs to explore the parameter space to collect relevant experience that is then used to adjust its prior model on the actual objective function.
However, while essential for the algorithm, the succession of sampled points might as well be very erratic:  probing consecutive points which fall far away in space from each other may pose machine-safety concerns, besides being unrealistic on real hardware where knobs are to be controlled.
As a dynamic system, we wish not to apply controls that are too different from the previous ones to avoid excessively stressing the laser.
Fig.~\ref{fig:BayesControls} shows the one-step absolute-value percentage variation distribution in the controls applied by BO in Ref.~\citenum{capuano2022laser} for each dispersion coefficient. 
Importantly, for all the $\psi_i$, a significant fraction of succeeding generated control signals vary drastically beyond practical and safe implementation on real hardware.
Moreover, Fig. \ref{fig:BayesControls_ActionRange} shows how BO controls are more than half of the times larger than 10\% of the size of the laser control bounds, which further confirms that a naive application of BO might be causing the controller to apply actions with a large magnitude fairly erratically.
As real-world function evaluations must be performed, we must also ensure that automatically generated controls respect basic requirements on the one-step relative change that the system can safely bear. 
Furthermore, using simulation-only solutions that are not specifically designed for transferability to the real world (\textit{i.e.,} optimizing a simulated version of the laser only) can result in poor performance when  knowledge of the system parameters is imprecise or when these they shift over time, leading to changes in the optimal solution $\psi^*$.

This work investigates a more general approach to laser pulse shape optimization which does not aim to produce a fixed set of control parameters $\psi$. 
Instead, it focuses on finding a policy to select control parameters given the system's current state and a goal to achieve.
Importantly, learning a control strategy rather than a fixed controller configuration improves adaptivity to changes in the experimental setup.
In this work, such control strategy is learned through Deep Reinforcement Learning (DRL), an advanced branch of machine learning that combines the principles of Reinforcement Learning (RL) and Deep Learning (DL) to address complex sequential decision making and control problems.

\newpage
\noindent The main contributions of this work are summarized as follows: 
\begin{itemize}
    \item We demonstrate that model-free DRL can be employed to optimally control HPL systems;
    \item We propose a DRL agent design integrating machine-safety constraints in its interactive learning strategy;
    \item We show that a physics-informed controller that only acts based on measurable physical quantities outperforms controllers acting based on temporal shapes.
\end{itemize}

Our extensive experimental evaluation demonstrates the ability of DRL to consistently optimize laser temporal shapes to nearly transform-limited pulses in less than 10 control timesteps starting from a random $\psi$ initialization.

\section{BACKGROUND}

\begin{figure}
    \centering
    \includegraphics[width=0.8\textwidth]{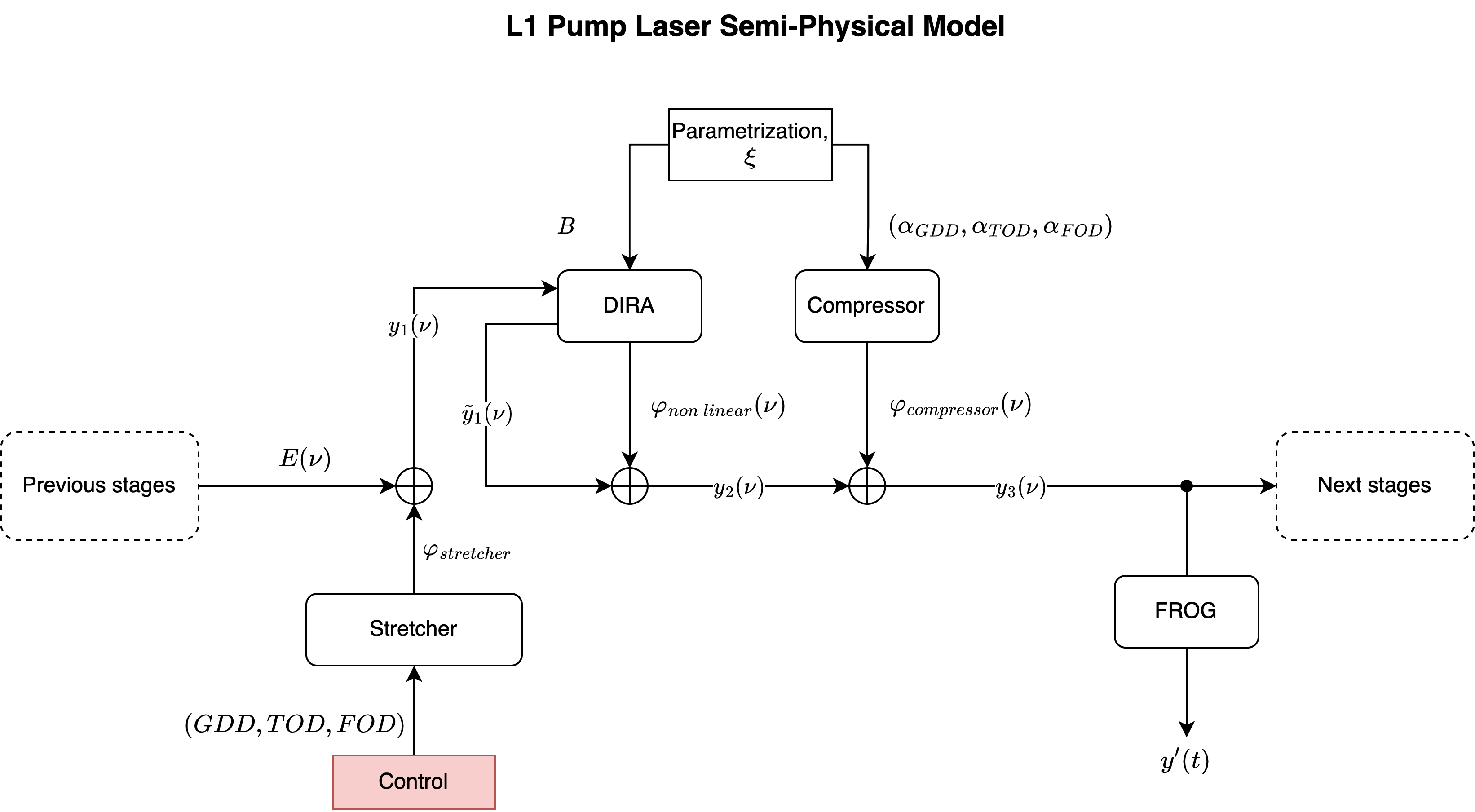}
    \caption{L1 Pump semi-physical model. The model has been built to simulate the process of phase accumulation in the L1 pump laser chain before Second-Harmonic Generation. 
    An input electric field $E(\nu)$ enters the chain from the left and undergoes a series of linear (stretcher and compressor) and non-linear (DIRA) phase accumulation processes. 
    Then, the corresponding temporal pulse shape is reconstructed through a FROG device and its second harmonic efficiency is computed using a proper measurement device.}
    \label{fig:L1SemiPhysical}
\end{figure}

\subsection{L1 pump semi-physical model}
This semi-physical model presented in Fig. \ref{fig:L1SemiPhysical} shows that our model contains a stretcher, a DIRA amplifier, and a compressor. 
Each stage in our model contributes to the overall pulse spectral phase. 
The stretcher imposes a spectral linear phase change, while DIRA introduces a nonlinear part of the spectral phase.
The linear DIRA contribution to the spectral phase is instead neglected. 
The dispersion coefficients of the compressor were measured beforehand and are kept constant throughout the experiments to simulate our best guess on the real-world experimental dynamics. 

Our model is parameterized by a non-linear phase accumulation term ($B$, standing for B-integral) and the compressor parameters ($\alpha_{\text{GDD}}$, $\alpha_{\text{TOD}}$, $\alpha_{\text{FOD}}$). 
The compressor parameters are well-known and do not typically change throughout operations. On the contrary, the value of B-integral cannot be precisely measured and we estimated it to be B $\simeq$ 2.
For the sake of notation simplicity, we refer to any given set of control parameters (GDD, TOD, FOD) as $\psi$ and to a generic laser parameterization as $\xi = (B,  \alpha_{\text{GDD}}, \alpha_{\text{TOD}}, \alpha_{\text{FOD}})$. 
We refer to our best estimate of the real-world parameterization as follows: 
\begin{equation*}
    \xi^* = 
    \begin{pmatrix}
    B^* \\
    \alpha^*_{\text{GDD}} \\
    \alpha^*_{\text{TOD}} \\
    \alpha^*_{\text{FOD}}
    \end{pmatrix}
    = 
    \begin{pmatrix}
    2 \\
    -267.422 \ ps^2 \\
    +2.384 \ ps^3 \\
    -954.89 \ ps^4
    \end{pmatrix}.
\end{equation*}

\subsection{Deep Reinforcement Learning}
In RL, an agent learns to make optimal decisions by interacting with an environment, receiving feedback in the form of rewards or penalties, and ultimately using it to adjust its actions. 
On the other hand, DL is a powerful technique for learning complex representations and function approximations using Deep Neural Networks. 
By integrating these techniques, DRL enables the development of agents capable of handling continuous state and action spaces and adapting to complex and dynamic environments.

RL involves learning to choose actions in a given situation to maximize a numerical reward signal~\cite{sutton2018reinforcement}.
The agent does not apply a predefined control strategy but has to leverage its experience to discover actions that yield the highest reward. 
Generally speaking, actions may affect not only the immediate reward but also the subsequent states and, therefore, all rewards in the considered time horizon.
The two key features distinguishing Reinforcement Learning from other types of learning are trial-and-error search and delayed reward, which make this technique applicable to a wide range of control tasks. 
\newline In our case, the goal is to gradually approach the optimal set of control parameters rather than converging to the single optimum value---as with black-box numerical methods.

In RL, the problem of learning from interactions how to solve a given task is formalized via Markov Decision Processes (MDPs). 
The learner (\textit{i.e.}, laser controller) is referred to as \textit{agent}, while whatever it interacts with is referred to as \textit{environment} (\textit{i.e.}, the L1 pump chain).
The agent-environment abstraction provides a simple interface to model the sequential interaction between the agent, selecting what actions to perform, and the environment, responding to the selected actions 
with a dynamic state transition and a reward, \textit{i.e.}, a numerical quantity encoding the utility of the action's effects on the system~\cite{sutton2018reinforcement}. 
Fig. \ref{fig:L1_AgentEnv} illustrates this sequential decision making process for the L1 pump laser.
This framework is used to reduce the control problem to a problem in which the agent has to learn how to maximize the reward it collects by choosing better actions over time.
In RL, one typically makes the assumption that the agent interacts with the environment at discrete time steps $t=0,1,2,3, \dots$. When the number of timesteps, the agent is allowed to perform is not upper-bounded, the MDP is classified as \textit{infinite-horizon}.
On the contrary, when timesteps are limited by a maximum value $T$, the MDP is classified as finite-horizon or \textit{episodic}. 

\noindent Every MDP $\mathcal M$ is defined by a tuple $\mathcal M = \{\mathcal S, \mathcal A, r, \mathcal D, \rho_0\}$, in which: 
\begin{itemize}
    \item $\mathcal S$ represents the state space, \textit{i.e.}, $s_{t} \in \mathcal S \ \forall t$. In those scenarios in which the agent has access to perfect information about the environment (\textit{fully observable} problems), the terms \textit{state} and \textit{observation} can be used interchangeably.
    \item $\mathcal A$ represents the action space, \textit{i.e.}, the space of all possible actions. That is, $a_t \in \mathcal A \ \forall t$.
    \item $r : \mathcal S \times \mathcal A \times \mathcal S \mapsto \mathbb R$ represents the reward function characteristic of the transition $(s_t, a_t, s_{t+1})$. As aforementioned, the agent's ultimate goal is the maximization of the cumulative reward value, often referred to as \textit{return}. 
    \item $\mathcal D$ describes the environment dynamics (forward dynamics) and represents the probability of the environment being in state $s_{t+1}$ having performed the action $a_t$ when in $s_t$. That is, $\mathcal D: \mathcal S \times \mathcal A \times \mathcal S \mapsto [0,1]$ with $\mathcal D(s_{t+1}, a_t, s_t) = \mathbb P(s_{t+1} \vert a_t, s_t)$. 
    \item $\rho_0$ describes the probability distribution over the initial state, namely $\rho_0= \mathbb P(s_0)$.
\end{itemize}

RL methods that estimate $\mathcal D$ by leveraging collected experience are classified as \textit{model-based}. 
In contrast, those without a forward dynamics model are referred to as \textit{model-free}.
Model-based algorithms are typically more sample-efficient (in the sense that they require fewer actual interactions) because the agent can learn to use its model of $\mathcal D$ to plan ahead its actions. 
However, this type of approach can only be applied when experience is a valuable source of information to reconstruct the actual dynamics (an example of this are all those cases in which a good prior on the transition model's stochastic process is available).
While being less sample-efficient, model-free approaches are instead applicable to a wider set of problems.

Let $\eta$ indicate the sequence of agent-environment interactions.
Then, any sequence of length $T$, often referred to as \textit{trajectory}, takes the form
\begin{equation}
    \eta = s_0, a_0, s_1, r_0, s_1, a_1, s_2, r_1, \dots, s_{T-1}, a_{T-1}, s_T, r_{T-1}.
\end{equation}
In light of the Markov property, it is trivial to prove that the probability mass of any trajectory equates to: 
\begin{equation}\label{eq:traj_prob}
    \mathbb P(\eta) = \mathbb P(s_0) \prod_{t=0}^{T-1} \mathbb P(s_{t+1} \vert s_t, a_t) \cdot \mathbb P(a_t \vert s_t),
\end{equation}
which clearly shows that the probability of the whole trajectory can be reduced to a series of piece-wise products between the probability of transitioning to state $s_{t+1}$ when performing $a_t$ in $s_t$ and the probability of selecting $a_t$ when in $s_t$. 
The probability of choosing $a_t$ when in $s_t$ is referred to as (stochastic) policy and is indicated by~$\pi(a_t \vert s_t)$. 
Stochastic policies are widely adopted in RL because they naturally embed randomicity, fostering the exploration of the action space conditioned on the current observation. 
Moreover, as the agent collects more and more experience it is not rare to observe that stochastic policies converge to deterministic ones anyways.  
In light of this, we can express the probability of observing one trajectory with respect to a given (stochastic) policy~$\pi$. 
This implies that Eq.~\ref{eq:traj_prob} could be re-written as: 
\begin{equation}\label{eq:traj_prob}
    \mathbb P_{\pi}(\eta) = \rho_0 \prod_{t=0}^{T-1} \mathcal D(s_{t+1}, a_{t}, s_t) \cdot \pi(a_t \vert s_t).
\end{equation}

Let $R(\eta)$ indicate the (undiscounted) cumulative reward over the trajectory $\eta$ (referred to as \textit{return} over $\eta$).
Then, the problem can be reduced to finding the policy maximizing the expected return, which in light of Eq.~\ref{eq:traj_prob} clearly depends on $\pi$. More formally, an optimal policy  satisfies:
\begin{equation}
    \pi^* = \arg \max_{\pi} \mathbb E_{\eta \sim \pi} \big[ R(\eta) \big].
\end{equation}
RL algorithms either optimize a parameterized version of the actual policy the agent is using to select actions (\textit{policy-based} algorithms) or the agent's estimate of the outcome of its own actions in terms of the return they yield (\textit{value-based} algorithms) with the goal of approximating said optimal policy.

\section{METHOD}
\subsection{DRL-based temporal shape optimization}
\begin{figure}[]
    \centering
    \includegraphics[width=.7\textwidth]{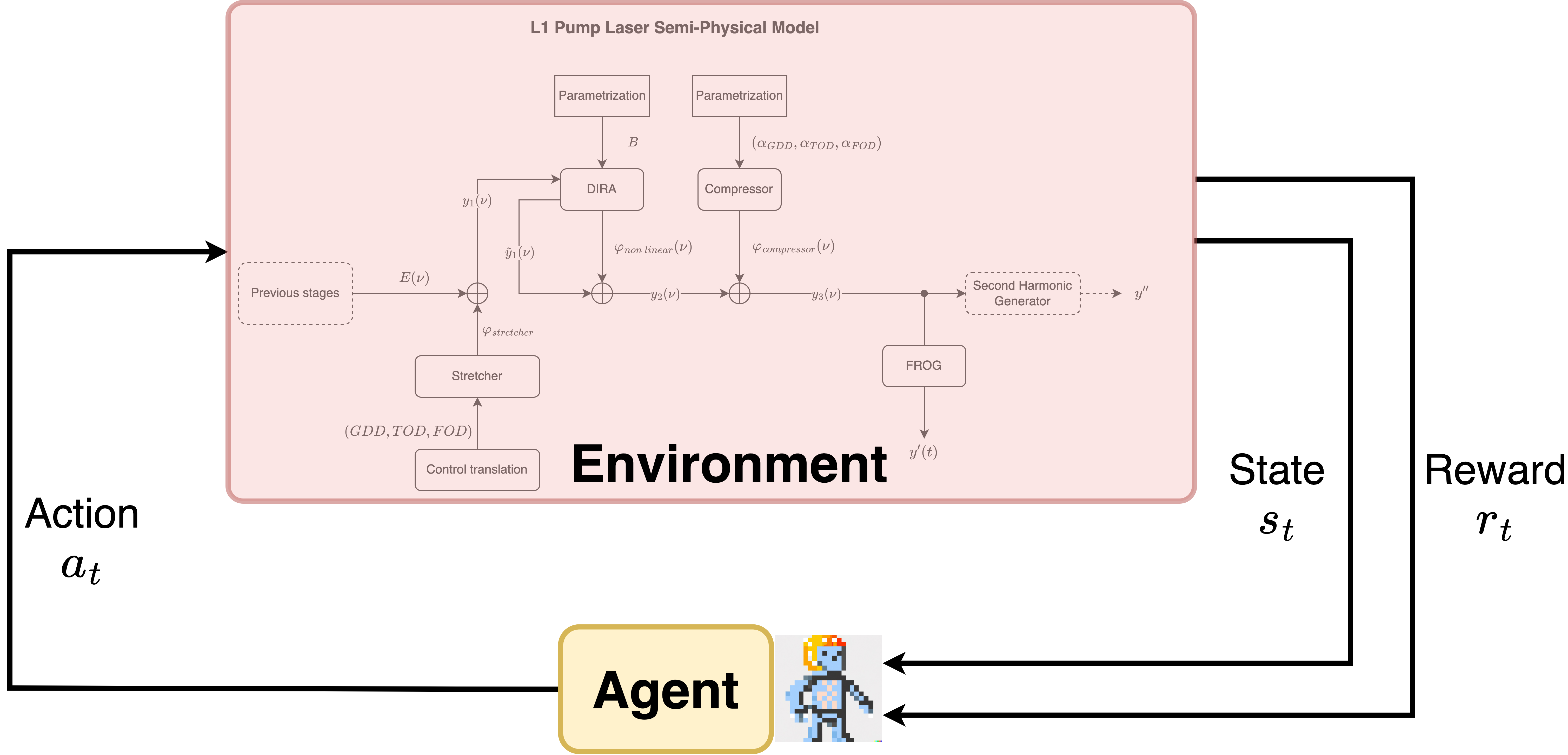}
    \caption{Agent-Environment interaction scheme for the L1 Pump Laser}
    \label{fig:L1_AgentEnv}
\end{figure}

As a flexible and powerful framework, DRL can be applied to a wide range of control problems. 
In this work, we present an application of DRL to laser pulse temporal shape optimization.
We only adopt model-free algorithms to keep our approach as general as possible. 
Fig.~\ref{fig:L1_AgentEnv} shows an application of the agent-environment paradigm to the control of the L1 Pump Laser. 
As it is possible to see, the agent selects the spectral phase parameters controlling the same semi-physical model, presented in Fig.~\ref{fig:L1SemiPhysical}, with the goal of maximizing the cumulative reward it can collect. 

\newpage
In this context, we define our two customized, fully-observable and episodic MDPs: 
\begin{enumerate}
    \item The first MDP (MDP-1) mainly stems from Ref. \citenum{capuano2022laser} and serves as a proof of concept that RL can indeed be used to control laser temporal shape.
    \item The second MDP (MDP-2) is a more physics-informed and sophisticated model that pursues the same goal as MDP-1 while improving the signal-to-noise ratio for the agent using a much more task-oriented reward signal.
\end{enumerate}
Both MDP-1 and MDP-2 share the same state and action space and  only differ by reward function definition and episode termination conditions.
Observations are defined as single control parameters $\psi$. 
This is justified considering the high degree of confidence with which these parameters are known in the real world, differently from what concerns the actual temporal shape. 
The latter can only be reconstructed using intermediate measurement devices (FROG), which  introduce a non-negligible amount of noise in the agent's perception of the environment.
This choice implies that, for what concerns the agent, the laser system only consists of the actual knobs of the Stretcher.
Given the nature of the problem, $\psi$ can only take values inside a feasible region, whose bounds we report (in SI units) in Tab. \ref{tab:feasible_region}. 

\begin{table}[t]
\centering
\caption{Feasible region for Stretcher control parameters}
\begin{tabular}{ccc}
\rowcolor[HTML]{EFEFEF} 
 & \textbf{Lower bound} & \textbf{Upper bound} \\
\textbf{GDD ($s^2$)} & 2.352e-22 & 2.996e-22 \\
\rowcolor[HTML]{EFEFEF} 
\textbf{TOD ($s^3$)} & -1.004e-34 & 9.556e-35 \\
\textbf{FOD ($s^4$)} & 4.774e-50 & 1.432e-49
\end{tabular}
\label{tab:feasible_region}
\end{table}

Actions are defined as deltas applied to the  set of control parameters. Thus, actions have the scope 
of modifying the currently applied control $\psi$ and, in particular, can only increase or decrease its magnitude.
Let~$\psi^{\text{range}}_i = \big(\psi^{\max}_{i} - \psi^{\min}_{i}\big) \ \forall i = 1, 2, 3$ be the range of the controls that can be applied on a stretcher.
To ensure machine safety, we constrain the agent to only perform actions with a magnitude up to 10\% of the feasible control range by limiting actions to $a_t(i) \in [-0.1 \psi^{\text{range}}_i, +0.1 \psi^{\text{range}}_i] \ \forall i = 1, 2, 3$.
This design choice ensures that the agent would be allowed to have a comprehensive sense of the actual controls applied (through our design of the state space) while performing actions concerned with the integrity of the system. 

Importantly, MDP-1 and MDP-2 differ in terms of reward function definition and episode termination conditions. 
Let us denote by $\tau(s_t)$ the pulse temporal shape corresponding to the control $s_t$ (that is, $\psi_t$) and by $\tau^*$ an arbitrary target temporal shape. 
Then, the reward function for MDP-1 is
\begin{equation}\label{eq:v1_reward}
    r(s_t, a_t, s_{t+1}) \equiv r(s_t) = c_1 - c_2 \cdot \frac{\mathcal L(\tau(s_t), \tau^*)}{\mathcal L_{\max}},
\end{equation}
with $c_1, c_2$ being two weighting coefficients and $\mathcal L$ a loss function measuring how dissimilar the current controlled temporal shape is with respect to the target one. 
In light of the results presented in Ref.~\citenum{capuano2022laser}, we adopt the peak-on-peak sum-L1 loss as $\mathcal L$, that is 
\begin{equation*}
    \mathcal{L}_1(\tau(s_t), \tau^*) = \Vert \tau_{\text{shift}}(s_t) - \tau^* \Vert_1,
\end{equation*}
where $\tau_{\text{shift}}(s_t)$ indicates a time-shifted version of $\tau(s_t)$ such that the controlled and target pulse peak in the same instant. 
Manually shifting the controlled pulse in time before computing the loss value is justified by the fact that none of the control parameters influence the temporal position of the pulse peak, while they all influence the pulse's shape.
The two coefficients $c_1$ and $c_2$ scale down the reward signal to make it more suitable for neural networks.
Moreover, $c_1$ serves as a \textit{healthy reward} term, whose primary goal is to encourage the agent to reach the end of the episode. 
Since at each timestep, the agent is also penalized with the loss value, the combined effect of the healthy and performance reward terms  pushes the agent towards reaching and maintaining pulses with small values of loss with respect to the target. 
In MDP-1, episodes have a maximal length of 50 timesteps and be terminated early whenever $\mathcal{L}_1(\tau(s_t), \tau^*) \geq \mathcal L_{\max} = 500$

We propose a more physics-informed approach for the design of MDP-2.
In particular, we aim to reward the agent for actions that increase the peak intensity and reduce the duration of FWHM. 
Therefore, we propose a reward of the form 
\begin{equation}\label{eq:v2_reward}
\begin{split}
    r(s_t, a_t, s_{t+1}) &\equiv r(s_t) = (2c_1)^2 + c_2 \min \bigg(\frac{0.1}{1-x_t} - 0.1, 7 \bigg) - 0.1 \cdot \text{FWHM}_t, \\
    &\quad x_t = \frac{PI_t}{PI_{TL}},
    \qquad \quad PI_t = \frac{2E}{\pi w_0^2 \int_{-\infty}^{+\infty} \tau(s_t) dt},
\end{split}
\end{equation}
with $c_1, c_2$ being two coefficients used to align the agent's perception of its performance and the goal of optimizing the actual pulse shape. 
$E$ denotes the pulse energy and $w_0$ the beam radius, equal to 220 mJ and 12 mm, respectively.
We employ the peak intensity of the TL pulse to scale down the peak intensity and positively reward the agent  when producing temporal pulse shapes with a peak intensity similar to TL's. 
At the same time, we discourage the agent from producing pulses with a large FWHM by using a penalty term directly depending on FWHM. 
In MDP-2, episodes have a maximum duration of 50 timesteps and are terminated early when $\text{FWHM}_t \geq \text{FWHM}_{\max} = 20 \ (ps)$. 
In contrast to the former case in MDP-2 we also penalize the agent with a large negative reward of $-20$ for terminating early. This is done for the sake of preventing the agent from developing a behavior in which episode termination is sought early on to reduce the overall penalization it would get otherwise.

For both environments, once the episode terminates, the initial state $s_0$ is sampled according to $\rho_0 \sim \mathcal N_3(\mu, \Sigma)$, with $\rho_0$ being a multivariate normal distribution with mean vector equal to the opposite of the compressor parameters, that is: $$\mu = (267.22 ps^2, -2.384 ps^3, 954.89 ps^4)$$ and variance-covariance matrix equal to $\Sigma = 0.1 I$, with $I$ being the identity matrix.

\subsection{Experimental setting}
Considering the large amount of experience that RL agents often need to accumulate, we train our agents on our custom 
simulated 
semi-physical model.
We evaluate three different RL algorithms on the two environment versions  defined previously (\textit{i.e.}, MDP-1 and MDP-2).
In our analysis, we include the on-policy policy-gradient-based algorithms PPO~\cite{schulman2017proximal} and TRPO~\cite{schulman2015trust} and the off-policy (policy-gradient-based) entropy-regularized algorithm SAC~\cite{haarnoja2018soft}. 
TRPO and PPO adopt mechanisms limiting the magnitude of policy updates, thus ensuring that policies remain close across consecutive iterations.
Instead, SAC employs a soft value function and entropy regularization to optimize policies while exploring diverse actions and operating on continuous action spaces.
We test the above algorithms  with 5 different random seeds and discount factors. Given the maximum episode length of 50 timesteps, we tested a range of discount factors to limit the time horizon that most affects the discounted return. Specifically, we tested $\gamma=0.9$, $\gamma=0.8$, and $\gamma=0.7$, corresponding to limiting the time horizon the agent is concerned with at decision time to approximately 10, 5, and 3 timesteps, respectively
Additionally, we experimented with various values of $c_1$ and $c_2$ for the two MDPs.
To train the agents, we use the Adam optimizer with a constant learning rate of 3e-4.
To ease policy training with function approximators, we linearly map $\psi_i$ in the range $ [0,1]$ for all $i=1,2,3$. 

To evaluate performance, we test all algorithms for 25 episodes and compute several metrics related to the terminal state of each episode. 
This evaluation strategy allows us to analyze agent performance in terms of its ability to reach and maintain a target temporal profile for most of an episode. 
Over 25 test episodes, we evaluate the last-timestep average $\mathcal L_1(\tau(s_T), \tau^*)$ and average cumulative reward for MDP-1. 
For MDP-2, we compute the last timestep average $\text{FWHM}_T$ and average peak intensity ratio $x_T$. 

The environments are implemented using the OpenAI Gym API~\cite{brockman2016openai}. We employed the Stable-Baselines3~\cite{stable-baselines3} implementation of the three RL algorithms using the PyTorch framework for DNNs. 
Each agent is solely trained on CPU, with 15 parallel environments (i.e. CPU threads) on an AMD Ryzen Threadripper 5995WX  desktop processor, taking approximately ~3 hours to complete 700k timesteps. The training process is periodically paused every 10K timesteps to perform 25 test episodes and log metrics of interest.

\section{RESULTS}

\begin{figure}
    \centering
    \begin{subfigure}[b]{0.45\textwidth}
        \centering
        \includegraphics[width=\textwidth]{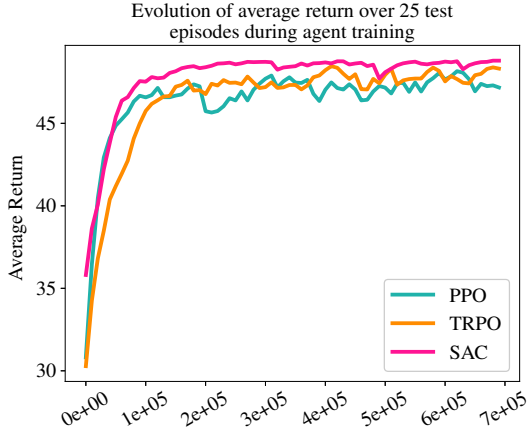}
        \caption{Average (over 5 seeds and 25 test episodes) evolution of the return collected during training.}
        \label{fig:v1_reward}
    \end{subfigure}
    \hfill
    \begin{subfigure}[b]{0.45\textwidth}
        \centering
        \includegraphics[width=\textwidth]{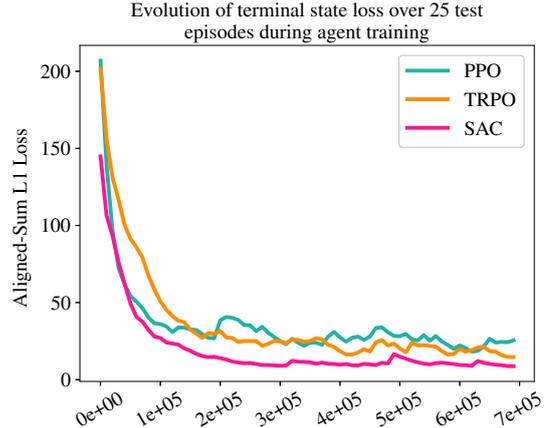}
        \caption{Average (over 5 seeds and 25 test episodes) evolution of the terminal $\mathcal L_1$ during training.}
        \label{fig:v1_loss}
    \end{subfigure}
    \caption{Evolution of the average return and terminal state $\mathcal L_1$ loss for the three algorithms. Here, are presented curves related to the best performing discount factor for all algorithms in light of Tab. \ref{tab:v1_results} and set of coefficients equal to $(c_1, c_2) = (0.1, 1)$. Plots are exponentially smoothed with coefficient $\alpha \simeq 0.18$.}
    \label{fig:v1_training_healthy_reward}
\end{figure}

\begin{figure}[t!]
    \centering
    \begin{subfigure}{0.45\textwidth}
        \centering
        \includegraphics[width=\textwidth]{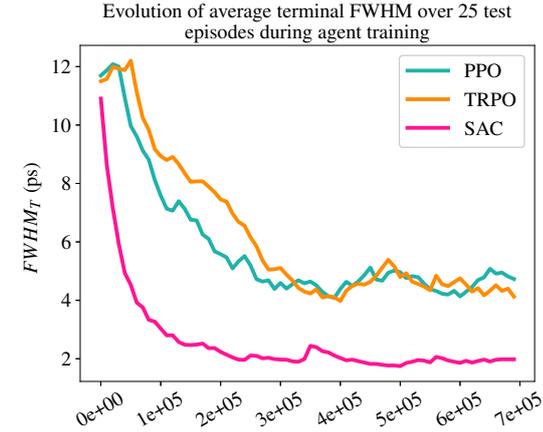}
        \caption{Average (over 5 seeds and 25 test episodes) evolution of the terminal state FWHM during training.}
        \label{fig:v2_FWHM}
    \end{subfigure}
    \hfill
    \begin{subfigure}{0.45\textwidth}
        \centering
        \includegraphics[width=\textwidth]{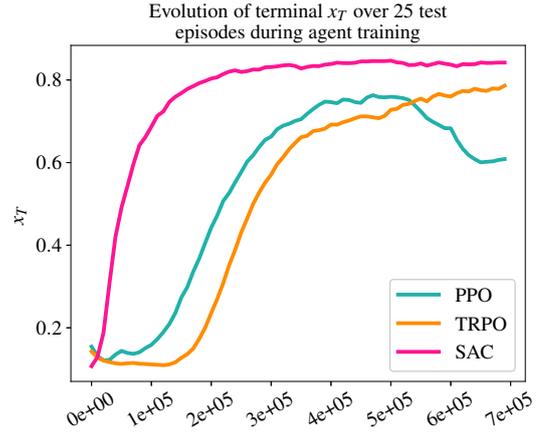}
        \caption{Average (over 5 seeds and 25 test episodes) evolution of $x_T$ during training.}
        \label{fig:v2_PI}
    \end{subfigure}
    \medskip
    \begin{subfigure}{0.45\textwidth}
        \centering
        \includegraphics[width=\textwidth]{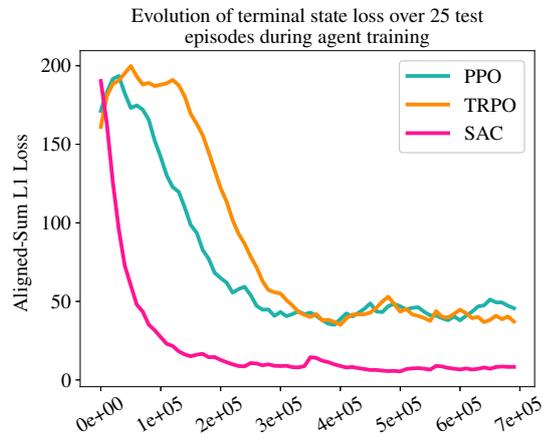}
        \caption{Average (over 5 seeds and 25 test episodes) evolution of the terminal state $\mathcal L_1$ during training.}
        \label{fig:v2_loss}
    \end{subfigure}
    \hfill
    \begin{subfigure}{0.45\textwidth}
        \centering
        \includegraphics[width=\textwidth]{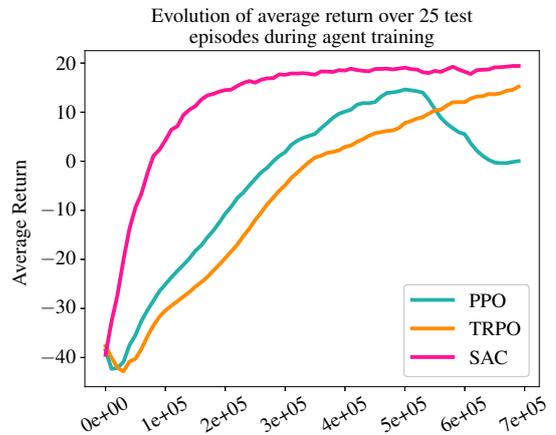}
        \caption{Average (over 5 seeds and 25 test episodes) evolution of the return collected during training.}
        \label{fig:v2_reward}
    \end{subfigure}
    \caption{Evolution of the average return and terminal state FWHM, Peak Intensity, and $\mathcal L_1$ loss for the three algorithms during training. Here, are presented curves related to the best-performing discount factor ($\gamma=0.7$) and the reward coefficients $(c_1, c_2) = (0, 1)$. Plots are exponentially smoothed with coefficient $\alpha \simeq 0.18$.}
    \label{fig:v2_training}
\end{figure}

Tab.~\ref{tab:v1_results} compares the performance of the three DRL algorithms on MDP-1 with different values for the discount factor~$\gamma$ and reward coefficients~$(c_1, c_2)$. 
In the interest of meaningful comparison, experiments are repeated five times with different random seeds to control for environment and agent stochasticity.
We report the median and inter-quantile range (IQR) statistics of the resulting performance metrics.

It can be clearly noticed that healthy reward terms have a positive impact on the final best result obtained, although not too drastic.  
Nevertheless, the fact that the best-performing configuration (SAC$+\gamma=0.7$) achieves comparable values of $\mathcal L_1$ between the two cases while the IQR when $c_1=0$ is almost twice as much as the case $c_1=0.1$ shows that rewarding the agent for being alive stabilizes its behavior at test time.

Fig. \ref{fig:v1_training_healthy_reward} shows the evolution of the two performance metrics previously defined during training. 
Notice that all algorithms learn how to reduce the value of the loss as the number of interactions  increases, even if SAC significantly outperforms TRPO and PPO on the task at hand.

Interestingly, Fig.~\ref{fig:v1_reward} suggests that the reward function might benefit from some further shaping. 
In particular, PPO achieves rewards similar to those achieved by the other algorithms for what concerns the average return, while having an average terminal loss value around 4 times larger than the one reached by TRPO, for instance. 
This indicates that the straight-forward reward function formulation  adopted for $v1$ might not be well-suited to help the agent discriminate between pulses, thus justifying the behavior exhibited in Fig. \ref{fig:v1_loss}. 

Fig.~\ref{fig:v2_training} shows the training curves for MDP-2, while Tab.~\ref{tab:v2_results}.
Since we adopt a penalty for early termination due to failure at pulse shaping, we set the healthy coefficient to $c_1 = 0$.
Remarkably, even though the agent is completely unaware of the laser characteristics and the very  concept of transform-limited pulse when faced with the task of maximizing the cumulative peak intensity while minimizing FWHM, it learned to control the temporal profile to maximize similarity with TL.
Once more, this demonstrates that transform-limited pulses achieve the best performance for the minimal duration in time and maximal peak intensity of laser pulses.
Furthermore, it is interesting to observe how, for both our reward functions, the agent performs best when it is less concerned with far-in-the-future consequences of its own actions
This is shown by the fact that better performances are obtained with smaller discount factors.

\begin{table}[]
\centering
\caption{Experimental results averaged over 25 test episodes on MDP-2 using 5 different random seeds. These results are solely related to the best-performing discount factor $\gamma = 0.7$. The best results are underlined. Each group of 5 experimental trials is presented via median value and inter-quantile range (IQR), between parentheses. Best results are underlined.}
\resizebox{\textwidth}{!}{%
\begin{tabular}{cccc}
\rowcolor[HTML]{EFEFEF} 
\textbf{Algorithm} & \textbf{Median Avg($\text{FWHM}_T$) (ps)  (IQR)} & \textbf{Median Avg($\mathcal L_1(s_T)$)  (IQR)} & \textbf{Median Avg($x_T$)  (IQR)} \\
PPO & 3.813 (1.719) & 31.93 (30.01) & 0.5689 (0.2083) \\
\rowcolor[HTML]{EFEFEF} 
TRPO & 2.581 (1.793) & 17.89 (32.49) & 0.8294 (0.07344) \\
{\ul \textit{SAC}} & {\ul \textit{1.615 (0.03964)}} & {\ul \textit{3.844 (0.43)}} & {\ul \textit{0.8559 (0.01244)}}
\end{tabular}%
}
\label{tab:v2_results}
\end{table}

Finally, Fig.~\ref{fig:v2_loss} shows that the best-performing agent does not only achieve quasi-TL pulses but is also able to maintain them over the entire length of the test episodes.
Note that this is a particularly challenging task that requires the agent to both learn to approach the target temporal shape starting from different initial conditions and then stabilize it.

\begin{figure}[t]
    \centering
    \includegraphics[width=\textwidth]{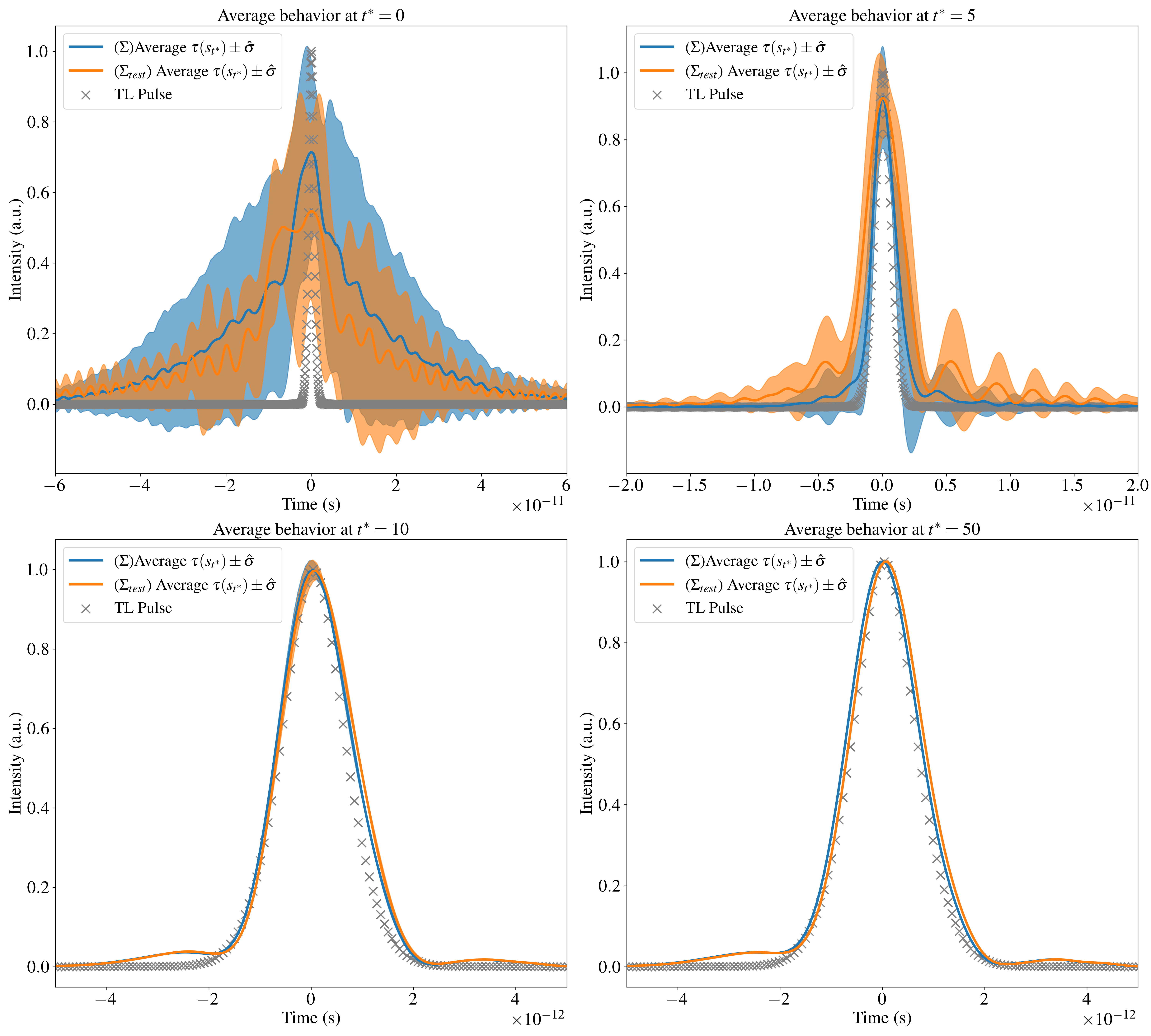}
    \caption{SAC-based agent trained on MDP-2 with $\gamma=0.7$ at test time. The plots show the mean temporal profile sampled at the 0, 5, 10, and 50 timesteps and the empirical standard deviation around the average $\hat{\sigma}$. Here we also confront two different initial state distributions (parametrized via $\Sigma$ and $\Sigma_{\text{test}}$, respectively), and show that the agent is capable of generalizing over random initial states.}
    \label{fig:sac_ood}
\end{figure}

Fig.~\ref{fig:sac_ood} shows that, in keeping with our expectations and the motivations of this work, the agent is capable of adapting its choices to the different control configurations it encounters. By sequentially updating $\psi$ the agent is able to reach nearly-TL profiles in less than 10 interactions with its environment, starting in a random $s_0 \sim \rho_0$. Moreover, we observe a decreasing empirical standard deviation around the average $\hat{\sigma}$ over time. This finding demonstrates the ability of the agent to both reach and maintain the desired target shape. Interestingly, Fig.~\ref{fig:sac_ood} shows an agent trained on MDP-2. \newline
Furthermore, we evaluated the robustness of our best-performing agent to different initialization conditions.
In particular, we sampled the starting configuration from a much higher variance $\Sigma_{\text{test}} = 50~\Sigma$, and report the results presented in Fig.~\ref{fig:sac_ood}. These results indicate that the agent is capable of reaching almost-TL shapes in the first 10 interactions, even in those settings in which the starting point is sampled from a significantly different distribution from the one seen during training. This appears to be particularly relevant, as it clearly shows that our agent is capable of generalizing over the starting state $s_0$.

\begin{table}[]
\centering
\caption{Experimental results averaged over 25 test episodes on MDP-1 using 5 different random seeds. This table indicates the median over the 5 episodes alongside the interquartile range (IQR), presented in parentheses for each experiment. Best results are underlined.}
\resizebox{\textwidth}{!}{%
\begin{tabular}{cccccc}
\textbf{} & \textbf{} & \multicolumn{2}{c}{\textbf{$(c_1, c_2) = (0.1, 1)$}} & \multicolumn{2}{c}{\textbf{$(c_1, c_2) = (0, 1)$}} \\
\rowcolor[HTML]{EFEFEF} 
\textbf{Algorithm} & \textbf{$\gamma$} & \textbf{Median Avg($\mathcal L(s_T)$)  (IQR)} & \textbf{Median Avg($R(\eta)$)  (IQR)} & \cellcolor[HTML]{EFEFEF}\textbf{Median Avg($\mathcal L(s_T)$)} & \cellcolor[HTML]{EFEFEF}\textbf{Median Avg($R(\eta)$)} \\
PPO & 0.7 & 30.01 (46.95) & 46.93 (3.727) & 10.53 (7.848) & 48.60 (0.783) \\
\rowcolor[HTML]{EFEFEF} 
PPO & 0.8 & 11.85 (13.97) & 48.34 (0.9436) & 10.58 (2.893) & 48.51 (0.235) \\
PPO & 0.9 & 23.30 (20.07) & 47.34 (2.273) & 41.78 (32.05) & 45.81 (2.000) \\
\rowcolor[HTML]{EFEFEF} 
TRPO & 0.7 & 6.147 (3.360) & 48.88 (0.2292) & 6.395 (2.800) & 48.96 (0.1799) \\
TRPO & 0.8 & 10.23 (55.49) & 48.65 (4.146) & 7.122 (19.67) & 48.88 (1.895) \\
\rowcolor[HTML]{EFEFEF} 
TRPO & 0.9 & 9.291 (26.88) & 48.49 (2.164) & 33.83 (18.06) & 46.78 (1.805) \\
{\ul \textit{SAC}} & {\ul \textit{0.7}} & {\ul \textit{6.014 (0.817)}} & {\ul \textit{48.95 (0.07063)}} & {\ul \textit{6.554 (1.926)}} & {\ul \textit{48.94 (0.0517)}} \\
\rowcolor[HTML]{EFEFEF} 
SAC & 0.8 & 8.819 (4.376) & 48.69 (0.4265) & 4.957 (6.436) & 48.97 (0.7563) \\
SAC & 0.9 & 19.14 (14.66) & 47.57 (1.052) & 18.07 (7.229) & 47.77 (0.6519)
\end{tabular}%
}
\label{tab:v1_results}
\end{table}

\section{CONCLUSIONS \& FUTURE WORKS}

In this work, we present a Deep Reinforcement Learning based approach to laser pulse temporal shape optimization.
This technique proved particularly promising for this task, as it allows us to obtain control strategies capable of gradually approaching desired optimal control rather than converging to static optimal solutions".
Moreover, it allows the optimization of the laser parameters even when the system is subject to complex nonlinear temporal dynamics, which are often unfeasible to model explicitly. 

We demonstrate that in just three hours of simulation-only interactions it is possible to train a DRL agent able to optimally control the spectral phase to reach a desired temporal shape. Importantly, this policy also obeys machine safety constraints by moving in bounded action steps.
Starting from random initial conditions centered around the compressor coefficients, the agent can effectively reach and maintain stable quasi-TL pulses, achieving pulses with FWHM of $\sim 1.6 ps$ in less than 10 interactions.

We plan to further expand our analysis by incorporating real-world data in future work. 
We will employ experimental observations to adjust our semi-physical model via inference and explore the application of DRL to laser optimization drawing inspiration from the Sim-to-Real paradigm in robot learning. 
Indeed, we believe there are various overlaps between Sim2Real in RL for robotics and the problem at hand. 
Since training robotics systems in the real world can be risky and costly, it is common to resort to training agents in a simulated environment. 
However, policies learned in simulation notoriously suffer from the reality gap---a discrepancy between simulated and real-world models---ultimately hindering policy transfer. 
To tackle this issue, several techniques have been developed to ensure policy transferability. 
As Sim-to-Real techniques are gaining more and more relevance in RL for robotics dynamical systems, we aim to investigate their applications to design an effective DRL-based self-tuning laser system in the real world.

\acknowledgments  
This study has been supported by the IMPULSE project which receives funding from the European Union Framework Programme for Research and Innovation Horizon 2020 under grant agreement No 871161. \newline
This study was also carried out within the FAIR - Future Artificial Intelligence Research and received funding from the European Union Next-GenerationEU (PIANO NAZIONALE DI RIPRESA E RESILIENZA (PNRR) – MISSIONE 4 COMPONENTE 2, INVESTIMENTO 1.3 – D.D. 1555 11/10/2022, PE00000013). This manuscript reflects only the authors’ views and opinions, neither the European Union nor the European Commission can be considered responsible for them.
G. T. also acknowledges the support by the EFORT Group in the context of the Italian National PhD in AI Programme.

\bibliography{report} 
\bibliographystyle{spiebib} 

\end{document}